\documentclass[tighten,apjl]{emulateapj}
\usepackage[]{times, graphicx}
\slugcomment{{\sc Accepted to Appear ApJ:} February 6, 2012}

\newcommand{\hinode}{{\em Hinode{}}}

\newcommand{\sdo}{{\em SDO{}}}
\newcommand{\pref}{\protect\ref}

\begin{document}

\shorttitle{Coronal Contraflow}
\shortauthors{Scott W. McIntosh et al.}
\title{On The Doppler Velocity of Emission Line Profiles Formed in the ``Coronal Contraflow''  that is the Chromosphere-Corona Mass Cycle}
\author{Scott W. McIntosh\altaffilmark{1,$\dagger$}, Hui Tian\altaffilmark{1}, Marybeth Sechler\altaffilmark{1}, Bart De Pontieu\altaffilmark{2}} 
\altaffiltext{1}{High Altitude Observatory, National Center for Atmospheric Research, P.O. Box 3000, Boulder, CO 80307}
\altaffiltext{2}{Lockheed Martin Solar and Astrophysics Lab, 3251 Hanover St., Org. ADBS, Bldg. 252, Palo Alto, CA  94304}
\altaffiltext{$\dagger$}{Movies supporting the figures (and a few extras) can be found at {\small \url{http://download.hao.ucar.edu/pub/mscott/papers/Contraflow/}}}

\begin{abstract}
This analysis begins to explore the complex chromosphere-corona mass cycle using a blend of imaging and spectroscopic diagnostics. Single Gaussian fits to hot emission line profiles (formed above 1MK) at the base of coronal loop structures indicate material blue-shifts of 5-10km/s while cool emission line profiles (formed below 1MK) yield red-shifts of a similar magnitude \-- indicating, to zeroth order, that a temperature-dependent bifurcating flow exists on coronal structures. Image sequences of the same region reveal weakly emitting upward propagating disturbances in both hot and cool emission with apparent speeds of 50-150km/s. Spectroscopic observations indicate that these propagating disturbances produce a weak emission component in the blue wing at commensurate speed, but that they contribute only a few percent to the (ensemble) emission line profile in a single spatio-temporal resolution element. Subsequent analysis of imaging data shows material ``draining'' slowly ($\sim$10km/s) out of the corona, but {\em only} in the cooler passbands. We interpret the draining as the return-flow of coronal material at the end of the complex chromosphere-corona mass cycle. Further, we suggest that the efficient radiative cooling of the draining material produces a significant contribution to the red wing of cool emission lines that is ultimately responsible for their systematic red-shift as derived from a single Gaussian fit when compared to those formed in hotter (conductively dominated) domains. The presence of counter-streaming flows complicates the line profiles, their interpretation, and asymmetry diagnoses, but allows a different physical picture of the lower corona to develop.
\end{abstract}


\section{Introduction}
Recent research has begun to explore the high-velocity mass injection from the chromosphere into the corona \citep[e.g.,][]{DePontieu2009, McIntosh2009a, DePontieu2011}. These episodic heating events are seen in association with a subset of spicules that are rooted in (predominantly) unipolar plage and network regions \citep[dubbed ``Type-II'' spicules by][]{DePontieu2007}. This investigation has been accompanied by a parallel effort to understand the complexity of the resulting ultraviolet (UV) and extreme-ultraviolet (EUV) emission line profiles in light of the quasi-periodic mass heating and injection taking place \citep[][]{McIntosh2010a, DePontieu2010b, 2010A&A...521A..51P, 2011ApJ...732...84M, Tian2011b, Tian2011c}. To date though, little has been made of the important, and inevitable consequence of the mass injection process \-- how and when the mass returns to the lower atmosphere from the corona and what is its signature? 

It has been well observed that the UV and EUV emission lines of the quiescent solar transition region (TR, formed below $\sim$0.8MK) ubiquitously exhibit a strong redshift \cite[e.g.,][]{Pneuman1978,Mariska1992,1997SoPh..175..349B,1998ApJS..114..151C,Peter1999,Curdt2008,2011arXiv1109.4493D}. It has been pointed out that the magnitude of the redshift is typically a few km/s stronger in the magnetic network versus inter-network regions \citep[similarly in coronal hole network, e.g.,][]{McIntosh2007} indicating that the presence of strong magnetic field, as well as the nature of the global magnetic topology, impacts the measured Doppler velocity. The TR redshifts are also present in active regions (ARs), with a magnitude typically larger than that in the quiet Sun \citep[e.g.,][]{Teriaca1999a,Teriaca1999b,Marsch2004,Marsch2008,DelZanna2008}. Recent observations with \hinode/EIS \citep[][]{Culhane2007} have revisited the ``cool plasma redshift,'' juxtaposing it with the systematic blue-shift of hotter lines (formed above $\sim$1MK) observed at the same location \citep[][]{2011ApJ...730...37U, 2011ApJ...727...58W, 2011arXiv1107.2362Y, 2011arXiv1107.1993K,2011ApJ...730...37U}. These observations have all highlighted the change in sign of the Doppler shift at $\sim$1MK.

Using a combination of \hinode/EIS emission line spectra and image sequences from the {\em Solar Dynamics Observatory} (\sdo) Atmospheric Imaging Assembly \citep[AIA;][]{2011SoPh..tmp..106L} we explore the spatial, spectral, and temporal behavior of the hot and cool emission of an active region. In the following sections we discuss our observations and analysis, that brings together line profile asymmetry and image analysis techniques, before we discuss the implications of the motions observed. Our observations highlight the complexity of the emission in the corona at temperatures below $\sim$1MK and suggest that we are simultaneously observing the heating and cooling phases of the chromosphere-corona mass cycle. 

\begin{figure*}
\plotone{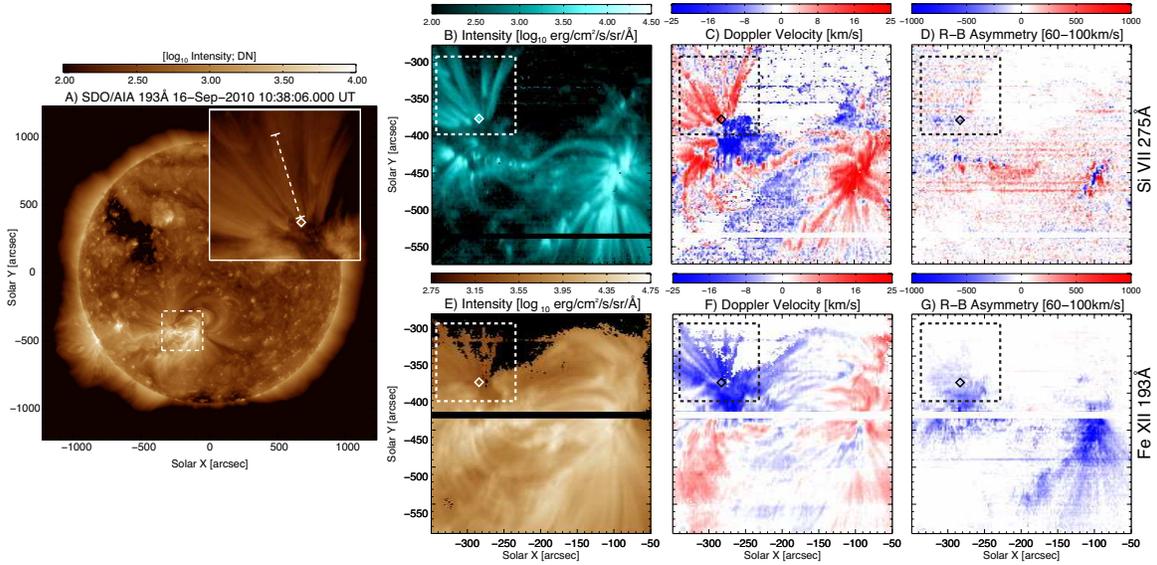}
\caption{The full AIA 193\AA{} field-of view (panel A) with a square dashed region showing the \hinode{} field of view - the upper left quadrant of which is shown inset. The inset also shows the location of the space-time plots studied in Figs.~\pref{f3}, and Figs.~\pref{f4}. Panels B through D, and E through G, show the line intensity, Doppler velocity, and 60-100~km/s RB in the \ion{Si}{7} 275\AA{} and \ion{Fe}{12} 193\AA{} emission lines respectively. The diamonds ($\diamond$) shows the sample location for Fig.~\pref{f2}. The electronic edition of the journal has supporting movies of the inset region. \label{f1}}
\end{figure*}

\section{Observations}
We analyze observations of AIA and EIS from September 16, 2010 taken between 10:38 and 11:57UT. The EIS sequence has an exposure time of 60s per slit position and the data were reduced using the ${\tt eis\_prep.pro}$ routine. The effects of slit tilt and orbital variation (thermal drift) were estimated by using the SSW routine {\it eis\_wave\_corr.pro} and removed from the data. The AIA images are delivered to the user with all appropriate corrections (flat-fielding, dark correction, etc) performed, they have a fixed cadence of 12s, and exposure times of 2.9, 2.0, and 2.9 seconds for the 131, 171 and 193\AA{} channels respectively.

\begin{figure}
\plotone{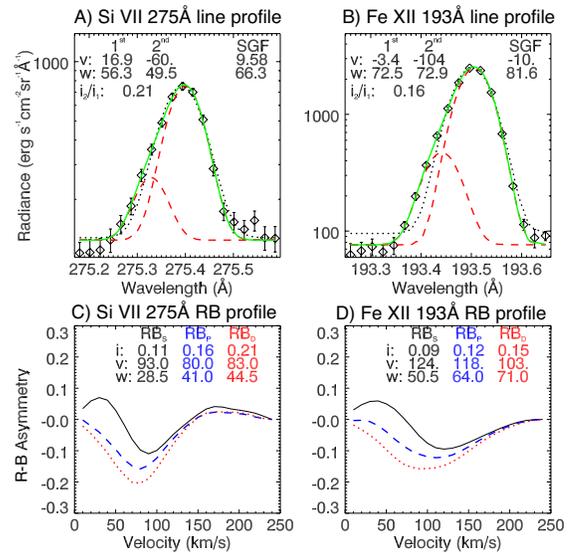}
\caption{Comparing sample single-pixel emission line profiles in \ion{Si}{7} 275\AA{} (A) and \ion{Fe}{12} 193\AA{} (B) from the location indicated in Fig.~\pref{f1} (error bars are given by the EIS reduction package). Each plot shows the SGF (dotted line) and DGF (green solid line) to the profile driven by the RB analysis (bottom row). Three variants of the RB analysis are shown in the bottom row \citep[][]{Tian2011b}, the standard method (RB$_{\mathrm S}$; solid line), the peak emission determined RB (RB$_{\mathrm P}$; dashed line), and the double fit determined RB (RB$_{\mathrm D}$; dotted line). \label{f2}}
\end{figure}

Figure~\pref{f1} provides context for this investigation. The AIA 193\AA{} field-of view (FOV) is shown in Panel A with a square dashed region indicating the EIS spectroheliogram FOV. We also show (inset) the upper left quadrant of the EIS FOV as a region where we will perform more detailed study. Panels B through D, and E through G, show the line intensity, Doppler velocity, and a measure of the line profile asymmetry \citep[``RB'';][]{DePontieu2009,Tian2011b} between 60 and 100~km/s for \ion{Si}{7} 275\AA{} and \ion{Fe}{12} 193\AA{} emission lines respectively. The diamond ($\diamond$) shows the location of the sample profiles extracted for Fig.~\pref{f2}. The RB diagnostic is the difference between the blue and red wing emission of a line profile following a single Gaussian fit to establish the line center and is discussed in detail by \citet{2011ApJ...732...84M} and \citet{Tian2011b}. The degree to which the line profile is asymmetric to the blue or red is reflected in the depth of color in the RB panels.

The electronic version of the journal has two movies of the inset region in the AIA 131 (predominantly emission of \ion{Fe}{8}), 171 (\ion{Fe}{9}), and 193\AA{} (\ion{Fe}{12}) channels. The individual images of the inset region have been corrected for solar rotation, co-aligned using a Fourier cross-correlation, and shifted to produce timeseries with an effective\footnote{The movies have one (five frames co-added) and five (twenty-five frames) minute cadence} exposure of 10s and 60s ensuring high signal-to-noise (S/N) in the three passbands, especially 131\AA{}. The purpose of the longer exposure will be evident below. The inset also shows the location of the space-time plots studied in Figs.~\pref{f3}, and~\pref{f4}.

\section{Analysis and results}
Due to the lack of cold chromospheric lines in the EIS spectra, it is not possible to perform absolute velocity calibration. Thus, we simply assume zero shift of the line profile averaged over the entire observation region and calculate the Doppler shift for each line profile. We immediately note (comparing panels C and F of Fig.~\pref{f1}) the presence of red-shifted \ion{Si}{7} emission and blue-shifted \ion{Fe}{12} emission on the same coronal structure (where we use the term ``structure'' to define a collection of magnetic field lines inside a spatial resolution element that have a consistent orientation). These oppositely directed line-of-sight (LOS) Doppler velocities determined by a single Gaussian fit (SGF) are of order 10km/s. The change from SGF red shift of the TR line to blue shift of the coronal line is very consistent with previous results \citep{2011ApJ...730...37U, 2011ApJ...727...58W, 2011arXiv1107.2362Y}. From panels D and G we see that both emission lines exhibit blue wing asymmetry in the 60-100km/s range at the roots of the loop system although we note that those for \ion{Si}{7} are considerably weaker. So, both lines display the presence of weak, high-velocity, blue asymmetries at the same location as the \ion{Si}{7} downflow and \ion{Fe}{12} upflow, as noted previously by \citet{McIntosh2009a} and \citet{Tian2011b}.

In Fig.~\pref{f2} we isolate sample line profiles at (-285\arcsec, -387\arcsec) the base of the inset loop system (identified by the $\diamond$ in Fig.~1) to illustrate the presence and correspondence of the asymmetries. Following the discussion of \citet{Tian2011b} (and see their Figs.~4, 9, and 17) we study the properties of the two line profiles. For \ion{Si}{7} and \ion{Fe}{12} (panels A and B respectively) we compare the single (SGF; dotted line) and double Gaussian fit (DGF; green solid line with individual components shown as red dashed lines) where the width and velocity position of the second component of the latter is constrained by the properties of their RB analyses (panels C and D). Again, we see that the SGF LOS Doppler velocities of these lines are of order 10km/s to the red and blue respectively. 

Next, we use the ``standard'' RB analysis \citep[RB$_{\mathrm S}$, i.e. that introduced by][using the fitted center of the SGF; black solid line]{DePontieu2009}, ``peak''  and  ``double'' RB analyses measures \citep[RB$_{\mathrm P}$ and RB$_{\mathrm D}$ were introduced by][as subtle variants of the method that use the velocity of the peak intensity \-- blue dashed line \-- and position of the dominant emission component \-- red dotted line as the reference point from which the asymmetry is computed respectively]{Tian2011b} to estimate the degree of asymmetry in the profiles. As pointed out by \citet{2011ApJ...732...84M} and \citet{Tian2011b}, the RB technique can not accurately resolve the secondary component if its velocity is smaller than the line width of the primary component. However, a small velocity of the secondary component usually leads to a negligible asymmetry of the total emission profile. Thus, the significant profile asymmetries shown in Fig.~\pref{f2} rule out this possibility. The appearance of the blueward asymmetries by spectroscopic blends in the blue wings of the line profiles can also be ruled out, since these blueward asymmetries are clearly present in not only one line, but in all strong and relatively clean lines in the EIS spectrum \citep[e.g.,][]{DePontieu2009,McIntosh2009a,Tian2011b,Bryans2010}. The large instrumental width of EIS certainly impacts the accuracy of the derived speed of the secondary component. However, joint imaging and spectroscopic observations presented in both this paper and \citet{Tian2011b} clearly demonstrate that the speed is indeed of the order of 100 km/s. From Fig.~\pref{f2} we note different LOS velocity offsets for the second components of the order of 85 and 115km/s respectively which, at first glance, imply a velocity difference with temperature \citep[also noticed by][]{Tian2011b}, but defer discussion to later. So, following the recipe of \citet{Tian2011b} we seed the DGFs with the RB$_{P}$ profile as well as the original SGF \-- the intensity of both components are free to vary but their velocity and width are constrained by the SGF velocity and width combined with the velocity offset and width of the RB$_{P}$ profile\footnote{The position and width of each component can vary by $\pm$22.3m\AA{} (25km/s at 275\AA{}) around their estimated values.}. From the DGF we see that the relative magnitudes of the second emission components are $\sim$0.2 and the LOS velocity offsets are 80 (\ion{Si}{7}) and 105km/s (\ion{Fe}{12}). {\em The DGF indicates that there are low amplitude high-speed blue asymmetries present in both lines at the roots of the active region where the SGFs indicate oppositely signed Doppler velocities.}

\begin{figure*}
\plotone{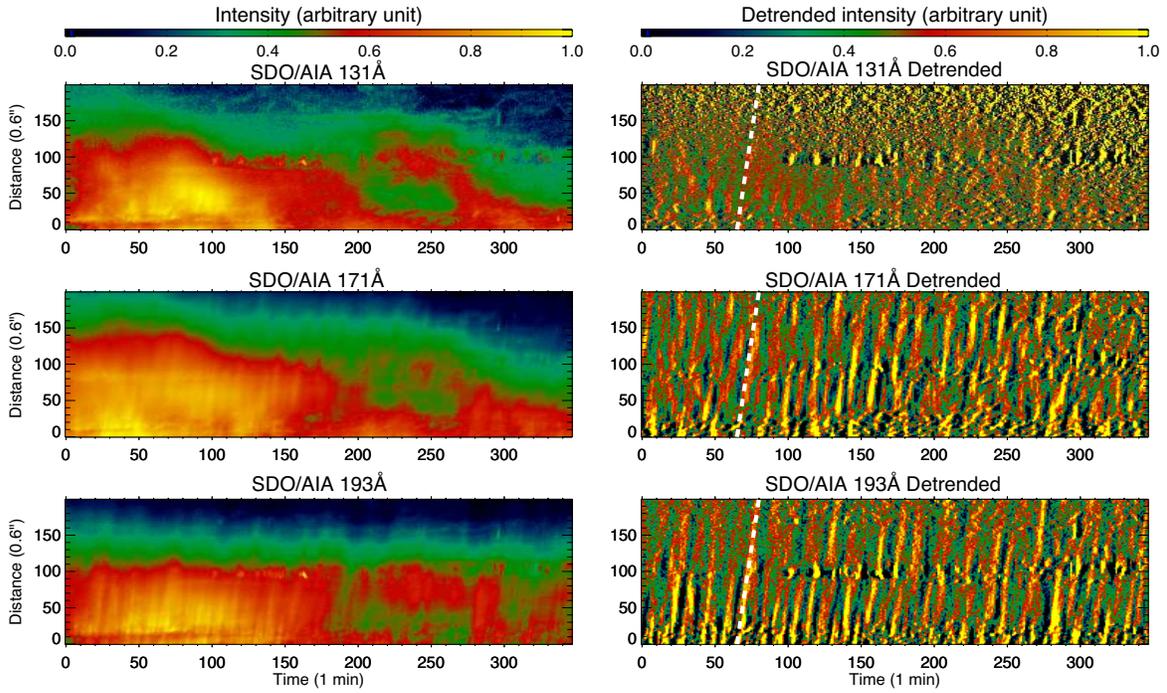}
\caption{Space-time plots of the intensity signal in the SDO/AIA 131, 171, and 193\AA{} from top to bottom along the path identified in Fig.~\pref{f1}. The data have been co-added to create an effective exposure time of 10s. The left column of panels show the variation of the raw intensity signal while the right column of panels show the 10 time step de-trended intensity signal to highlight the motion observed. The inclined white dashed lines indicates an apparent upward velocity of 96km/s. \label{f3}}
\end{figure*}

\begin{figure*}
\plotone{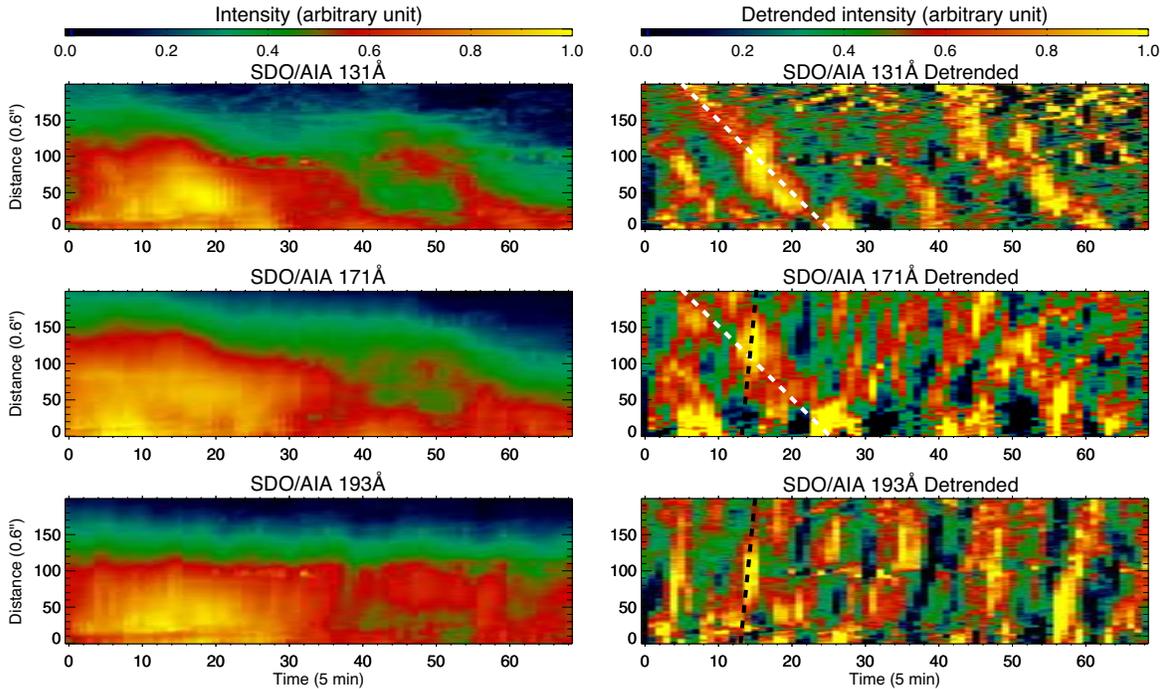}
\caption{Same as Fig.~\pref{f3} where the data have been co-added to create an effective exposure time of 60s. The inclined white dashed lines shown on the top two panels indicate an apparent downward velocity of 14km/s while the black dashed lines the 96km/s apparent velocity of Fig.~\pref{f3}. \label{f4}}
\end{figure*}

Now, we use the AIA 131, 171, and 193\AA{} channels to study the space-time evolution along the same coronal loop structure (dashed line if Fig.~\pref{f1}) where two of the passbands have similar formation temperatures (\ion{Si}{7} \-- 131\AA; \ion{Fe}{12} \-- 193\AA) and the 171\AA{} channel, is formed roughly in-between the two. We note that previous space-time analysis of similar coronal structures with TRACE, STEREO/EUVI, and AIA \citep[e.g.,][]{McIntosh2009a, DePontieu2011, Tian2011a, Tian2011b} have shown remarkably consistent plane-of-the-sky (POS) velocities (or apparent motions) measured in ``propagating coronal disturbances'' (PCDs) across a range of temperatures. However, we should note that these PCDs cannot be uniquely designated as rapid upflow events or upwardly propagating slow-mode MHD waves from imaging diagnostics alone \citep[e.g.,][]{DePontieu2010b,Tian2011c,Nishizuka2011}.

PCDs are clearly visible in the supporting 60s cadence AIA movies. Performing the space-time analysis of \citet{Tian2011b} we see the evolution of the PCDs along the coronal loop for the three AIA channels (131, 171, and 193\AA{}) in Fig.~\pref{f3} from top to bottom. The left panels show the raw intensity variation with time while the right panels show the signal following the removal of a 10 time-step running mean from the space-time plot. Clearly, the 131\AA{} signal is significantly noisier than the other two channels, but there is consistency in the left-right upwardly inclined stripes visible in the plots indicating the presence of PCDs on the structure at a range of temperatures as shown in \citet{Tian2011b}. The PCDs have a POS velocity of 96 ($\pm$10) km/s and a reference line is shown on all three panels to indicate this. The fact that the PCD speed shows no obvious temperature dependence seems to be very different from the increase of upflow speed with temperature as inferred from EIS line profiles. Such an inconsistency will be discussed in below. We note that the fact of redshifts in cool emission lines (e.g., \ion{Ne}{8}) and upflows in imaging observations at similar temperatures (e.g., TRACE 171 \AA{} passband) has already been pointed by \citet{Winebarger2002} and \citet{Schrijver1999}. So far no satisfactory explanation for this phenomenon has been provided. In the following we will explain it in the context of the mass cycling between the chromosphere and corona. 

Using a potential field extrapolation we estimate the inclination of the coronal structure at 55$\degr$ with respect to the LOS allowing us to consolidate the POS PCD motion and LOS blue wing EIS asymmetries \citep[see the discussion in Sect. 4.3 of][]{Tian2011b}. Therefore, their common rooting, commensurate velocities, as well as the consistency of the PCD velocities across temperature, are sufficient evidence for us to interpret the observed PCDs as the signature of rapid chromospheric mass transport at the base of the active region loops \citep[][]{DePontieu2011}. {\em Therefore, from this point on, we will assume that the observed PCDs are the signature of rapid upflow events originating at the base of the coronal structures, although we acknowledge that they may take on a different character away from their origin.}

Repeating the space-time analysis for the image sequence with an effective exposure time of 60s (the same exposure time as EIS) we see, in Fig.~\pref{f4}, a profound difference in the patterning visible. The 131\AA{} channel space-time plot is dominated by stripes pointing in the other direction - a reference speed of 14km/s is shown as a white dashed line - indicating an apparent downward motion of material along the structure. {\em There is no longer a clear signature of upward PCDs in the cool channel.} The 193\AA{} channel still shows upward PCDs but the temporal averaging has slowed them \-- the black dashed line shows 96km/s \-- with stripes in the space-time plot that are slightly more inclined. The signal in the 171\AA{} channel appears to be a mix of upward and downward apparent motions where the 96km/s upward (black) and 14km/s downward (white) lines are shown for reference. Studying the associated (long effective exposure) supporting movie we can easily discern the prevalence of slow downward motions in 131\AA{}, and fast upward motions in 193\AA{} with a mixture of the two in 171\AA{}.

\begin{figure}
\plotone{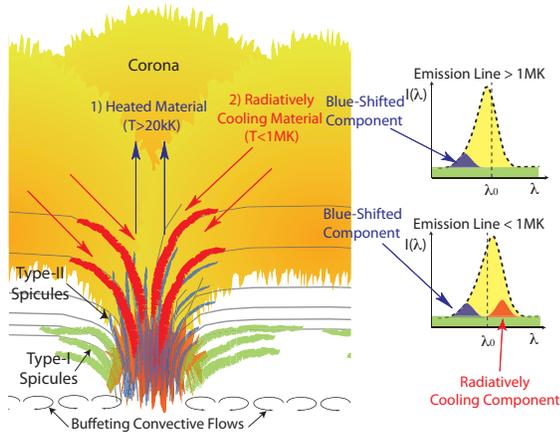}
\caption{Cartoon to illustrate the impact of the chromosphere-corona mass cycle on the line profiles observed. \label{f5}}
\end{figure}

\section{Discussion}
From the above analysis it is clear that we have to be incredibly careful when interpreting the data from spectrographs and broadband imagers at all times, but particularly for emission formed at temperatures below $\sim$1MK\footnote{We consider the regions studied, and their PCD/RB signatures to be very typical. This perspective is formed by studying many such loop systems with SDO/AIA and Hinode/EIS.}. 
There are co-spatial upflows (of order 100km/s) and downflows (of order 10km/s) on the coronal structures depending on the temperature of the diagnostic studied. The appearance of this ``contraflow'' of up- and down-flowing plasma streams on the same structure is contrary to the analysis of \citet{2011ApJ...730...37U} and \citet{2011ApJ...727...58W}, where the apparent lack of a blue wing asymmetry and oppositely signed (SGF) Doppler shift led the authors to assert that low temperature high-speed upflows were not present in the system. Such assertions must be made noting the impact that the viewing geometry of the structures (relative to the observer's LOS) that are critically important when considering the detection (or non-detection) of profile asymmetries in emission lines formed with large scale heights, such as the lines observed by EIS \citep[e.g.,][]{DePontieu2010b, 2011ApJ...732...84M, Tian2011b}. 

Evaluating the way in which cool emission lines are affected by the coronal contraflow is an essential avenue of future investigation (e.g., for the Interface Region Imaging Spectrograph) but is beyond the scope of this work. Further, the relative magnitude of the additional up- and down-flowing components to the emission of the cool lines is as a function of temperature is unclear, but careful differential analysis of coronal hole and quiet Sun plasmas and emission lines may hold vital diagnostic clues. For example, the UV emission lines typically show a reduced red-shift \citep[by as much as 5km/s, see e.g.,][]{Xia2004,McIntosh2007,McIntosh2010a,Tian2010} in coronal holes compared to the quiet Sun where the draining flow component may be significantly reduced (as there may be little, or no, corona above).

From what has preceded, it seems entirely reasonable that the process producing the slow drainage of material out of the corona is significantly contributing to the redshifts of single Gaussian fits to EIS EUV emission lines formed below $\sim$1MK, and is therefore also related to the strong TR redshifts observed in the quiet Sun \citep[e.g.,][]{Peter1999,2011arXiv1109.4493D}. We suppose that these draining flows relate to the cooling of coronal material and its final return to the lower atmosphere \citep[e.g.,][]{2003A&A...411..605M,2005A&A...436.1067M,2010ApJ...710L..39B,2010ApJ...716..154A}. The increased efficiency of the plasma to lose energy by radiation at temperatures below 1MK \citep[e.g.,][]{Mariska1992} will weight emission line profiles formed in cooling streams to radiate far more efficiently than the rapidly heating upflow that may, or may not, be in equilibrium. This combination of flows within the spatio-temporal resolution element of a spectrograph would tend to produce ensemble emission with net redshift in cool lines while the cumulative upflow of many PCDs visible in the hotter passbands, minus the radiatively cooling component where thermal conduction dominates the energy balance, can explain the apparent reversal in sign of the Doppler velocity behavior with temperature. That is, the cooling downflows are harder to see at the formation temperature of \ion{Fe}{12} where thermal conduction is dominating the cooling process, affecting the line profile less. Therefore, for emission lines in the neighborhood of magnetic flux elements, we have to worry about {\em three} components of the emission depending on the temperature of the diagnostic \-- the rapid, weak upflow generated in the lower atmosphere at the beginning of the mass cycle, the ``static'' coronal background emission that is possibly caused by gentle evaporation of cool material \citep[e.g.,][]{Patsourakos2006}, and a cooling/draining downflow. Figure~\pref{f5} attempts to show the possibilities pictorially \citep[cf. Fig.~5 of][]{DePontieu2009}. 


\begin{figure}
\plotone{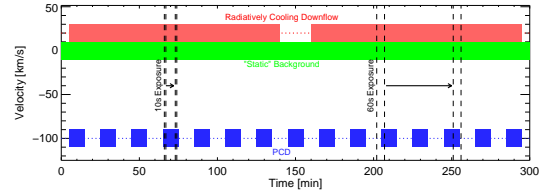}
\caption{Schematic illustration of the flow directions in the SDO/AIA 131\AA{} passband. The green, red and blue bars indicate the background, cooling downflow and rapid upflow components, respectively. The arrows indicate the sequences of temporal sampling. \label{f6}}
\end{figure}

Figure~\pref{f6} is a schematic illustration explaining the flow directions in the SDO/AIA 131\AA{} passband. We simply assume that there are two cooling downflows and 15 rapid upflows along a certain magnetic field line during an observation period of 300 minutes. The green, red and blue bars indicate the background, cooling downflow and rapid upflow components, respectively. The arrows indicate the direction of time-stepping. In the case of 10 s effective exposure (5 frames averaged, 1 minute cadence), a single upflow is well sampled from its beginning to its end. However, only a small portion of a single downflow is sampled during the lifetime of a single upflow. Thus, we can see the complete evolution of a single upflow but can hardly see any movement associated with the downflow in the space-time plot, as evident in Figure~\pref{f3}. In the case of 60 s effective exposure (25 frames averaged, 5 minute cadence), the exposure time is comparable to the lifetime of an upflow but much less than that of a downflow. In this case a single downflow can be well sampled from beginning to end. Although several upflows occur sequentially during the lifetime of a downflow the complete evolution of these upflows is not properly sampled. Moreover, for more than half of the time steps, the gaps between upflows are sampled, therefore the relative contribution of the upflow component to the total emission is reduced and the downflow contribution is thus enhanced as compared to the case of 10 s effective exposure. We propose that these two effects are responsible for the presence/absence of signatures of downflow/upflow in the 131\AA{} passband shown in Figure~\pref{f4}. 

Unfortunately, as we have demonstrated, the physical picture in the cool legs of coronal loop systems is far more complex than a simple SGF can possibly reveal. The presence of more than two emission components clearly has the potential to impact not only the SGF, but also our ``simple'' RB asymmetry analysis. As an example, consider the \ion{Si}{7} line which shows evidence of asymmetry, but produces systematically produces lower amplitude (and velocity) RB profiles (Fig.~\pref{f1}) even though the PCDs visible at those temperatures appear to be commensurate in strength and speed with their hotter counterparts. When the ensemble emission contains strong emission from downflowing material, a DGF scheme, or the RB analysis, is not adequate to describe the profiles observed \-- the weakly red-shifted, relatively bright, component can ``mask'' the appearance of the weaker upflow emission in the blue wing of the line \-- leading to reduced peak velocity and amplitude of inferred RB profiles and giving the appearance of a symmetric line profile \citep[e.g.,][]{2011ApJ...727...58W}. Therefore, we propose that the addition of the significant cooling downflow contribution to cool emission lines is probably responsible for the reduced upflow speed of \ion{Si}{7} as compared to \ion{Fe}{12}. However, we realize that new observations and better analysis techniques are needed to quantitatively evaluate this significant complication\footnote{As a matter of point, and based on the above investigation, we caution against the use of the SGF determined velocity-temperature relationship \citep[e.g.,][]{Peter1999} as a ``metric'' to test the output (physically sophisticated) numerical simulations \citep[e.g.,][]{2006ApJ...638.1086P} \-- for transition region like temperatures it is unclear what we are learning about the physics of these elaborate models using the coarse SGF representation of the complex emission.}.

Clearly, the coronal (and transition region) emission lines observed in the neighborhood of magnetic flux concentrations are not simple single component profiles. We have demonstrated that, with careful use of the imaging datasets available, it is possible to provide diagnostically useful information on the additional components of the emission that can exist in the complex interface of heated and cooling counter-streaming flows. Finally, we note that further observations and extensive modeling are required to accurately establish the contribution of the various emission components of the mass cycle. Understanding the balance of these components is critical to a better understanding of the interface between the lower and upper corona. Indeed, analysis such as that presented herein may allow us to investigate the presence of cool contaminants in the hotter \sdo/AIA basspands \citep[e.g., 211\AA{};][]{2011A&A...535A..46D}. If the hotter passbands are ``clean'' and show only emission from hot material then one should expect that no significant downflow to be present when summed image sequences are studied (cf. Fig.~\pref{f4}). Indeed, preliminary investigations with the AIA 211\AA{} passband would appear to indicate that there is indeed a (cool) contaminant, but that it's impact can only star being observed when images are summed to the level of those shown in Fig.~\pref{f4}. We expect that it may be possible to estimate the level of the contamination to the passband by adapting the analysis above, but that is left for future work (Tian et al. 2012 \-- in preparation).

\section{Conclusion}
In the vicinity of magnetic concentrations single Gaussian fits to emission line profiles are far from sufficient to describe the complex underlying physics of the interface between the chromosphere and corona. In particular, we suggest that the persistent downflows seen in cool emission are driven by the transition of the plasma into the radiative cooling domain and are a natural consequence of the mass-cycling between the chromosphere and corona representing the return of previously heated plasma to the lower atmosphere. The superposition of these flows in the ensemble emission of a spatio-temporal resolution element can provide a consistent explanation of the modulation observed in the single Gaussian fit line shift in the $\lesssim$1MK and $\gtrsim$1MK plasma regimes.

\begin{acknowledgements}
SWM, HT, and MS are supported by NSF ATM-0925177, NASA NNG06GC89G; BDP by NASA grants NNM07AA01C ({\em Hinode}), and NNG06GG79G. SWM and BDP are jointly supported by NASA grants NNX08AL22G, NNX08AH45G, and NNG08FD61C ({\em IRIS}). HT is also supported by the Advanced Study Program of NCAR. \hinode{} is a Japanese mission developed and launched by ISAS/JAXA, with NAOJ as a domestic partner and NASA and STFC (UK) as international partners. \sdo{} is a project of NASA. NCAR is sponsored by the National Science Foundation. 
\end{acknowledgements}

\end{document}